\documentclass[preprint,aps,prl,showpacs]{revtex4}
\usepackage{epsfig,graphics,amsmath}
\newcommand{\bea}{\begin{eqnarray}}
\newcommand{\eea}{\end{eqnarray}}
\newcommand{\be}{\begin{equation}}
\newcommand{\ee}{\end{equation}}

\begin{document}
\title{Probing Disordered Substrates by Imaging the
Adsorbate in its Fluid Phase}
\author{Ankush Sengupta\footnote{E-mail : ankush@bose.res.in}
and Surajit Sengupta\footnote{E-mail : surajit@bose.res.in}}
\affiliation{S.N. Bose National Centre for Basic Sciences,\\Block JD, 
Sector III, Salt Lake,\\ Kolkata 700 098,\\ India.}
\author{Gautam I. Menon\footnote{E-mail : menon@imsc.res.in}}
\affiliation{The Institute of Mathematical Sciences,\\ C.I.T. Campus,
Taramani, Chennai 600 113,\\ India.}

\begin{abstract}

Several recent imaging experiments access the equilibrium
density profiles of interacting particles confined to
a two-dimensional substrate.  When these particles are
in a fluid phase, we show that such data yields precise
information regarding substrate disorder as reflected
in one-point functions and two-point correlations of
the fluid.  Using Monte Carlo simulations and replica
generalizations of liquid state theories, we extract
unusual two-point correlations of time-averaged density
inhomogeneities induced by disorder. Distribution functions
such as these have not hitherto been measured but should
be experimentally accessible.

\end{abstract}
\date{\today}
\pagebreak
\maketitle

Place a perfect crystal in a disordered background deriving
from a large number of randomly placed, quenched point
pinning sites.  Allow the crystal to relax to its minimum
free energy state in this background. A qualitative
picture of structure in the state which results is
obtained by balancing the energy cost of distortions
in the displacement field with the energy gained from
accomodating locally to the pinning, a  competition
conventionally described via the paradigm of an elastic
manifold in a random medium \cite{larkin,halpin}.
Such a starting point, however, is clearly inadequate
when equilibrium correlations in the pure system do not
resemble those of a crystal or indeed of any system in
which particles of fixed connectivity are separated by
a mean distance from which deviations are energetically
penalized. The appropriate problem in this case is that of
determining the correlations of an {\em equilibrium} fluid
in a quenched random environment\cite{gau}. 
Such a description applies to a large number of experimental
systems.  Indeed, systems modelled as elastic manifolds at low
temperatures generically melt into fluid states at high
enough temperatures.

Equilibrium fluids in a quenched disordered background
exhibit disorder-averaged versions of the conventional
correlation functions of the pure system. In addition, new
classes of non-trivial correlations emerge \cite{gau}.
Modern imaging techniques enable direct microscopic
visualization of several two-dimensional situations
which may be mapped onto the problem of interest here:
these include colloidal particles in their fluid phase
on rough substrates\cite{kusner,kusner1,Zahn},
magnetic bubble domain arrays at large
effective temperatures\cite{seshadri1},
charge-density wave systems\cite{dai} and varied
phases of vortices in thin superconducting
films\cite{harada,moser,pal,oral,soibel1}. As we
show in this Letter, such experiments raise the
possibility of measuring novel correlation functions
associated with fluid systems in the presence of quenched
disorder. In addition, the topography of surface randomness
can be accessed from the images yielded by such methods,
even though it is the fluid (adsorbate) particles which
are being imaged and not the substrate directly.

This Letter describes a procedure for reconstituting a
disorder potential in microscopic detail from knowledge of
the time averaged density $\rho({\bf r})$ of the adsorbate
in the liquid state and its spatial correlations. To this
end, we also present accurate benchmarks for replica-based
liquid state theories for correlations in disordered
systems, comparing the predictions of these theories with
direct simulations.  Our proposals are illustrated using Monte
Carlo computer experiments on a model two-dimensional fluid
in a disordered background.  Figure~\ref{pot} exhibits
our main result: Fig.\ref{pot} (a) shows the random
substrate potential used in our Monte Carlo calculations,
while Fig. \ref{pot} (b) exhibits the reconstituted potential
obtained from the {\em induced} $\rho({\bf r})$,
using a replica-based liquid state theory
\cite{gau}. As these figures illustrate, to an
accuracy limited only by the computational effort required
for a well-averaged $\rho({\bf r})$, our calculation
reconstitutes the imposed disorder potential.

Our model system comprises $N = 780$ particles
confined to two dimensions and interacting {\it via}
a inverse-twelfth power pair potential. These particles
also interact with a quenched one-body disorder potential
$V_d({\bf r})$,  modelled as a zero-mean Gaussian random
field with specified variance and short-ranged spatial
correlations. The interaction part of the Hamiltonian
is thus: $H_{int} = \epsilon\sum_{i<j}(\frac{\rm
\sigma_{0}}{r_{ij}})^{12} + \sum_{i}\it{V_{d}}({\bf
r}_i)$. We set $\epsilon=1$ and ${\rm \sigma_{0}}=1$,
thus setting energy and length scales.  We use periodic
boundary conditions, benchmarking our calculations
in the pure limit against earlier work\cite{BGW}.

\begin{figure} 
\vskip 6.2 cm
\caption{(color-online) (a) Color plot of the 
disorder potential $\beta V_d(x,y)$ used in our Monte Carlo
simulation. The plot is a single realization of a Gaussian random 
potential of strength $\sigma^2 = .1$ and correlation 
length $\xi = .12 \sigma_0$. (b) The reconstituted potential
(see text) with the measured time averaged density $\rho({\bf r})$ 
(Fig. 2 (a)) as input. The correspondence between (a) and (b) 
is close but can be systematically improved by averaging 
$\rho({\bf r})$ for longer times. The $\sigma^2$ and $\xi$ 
obtained from (b) agree with the input values in (a) to within 
a few percent. 
}\label{pot}
\end{figure}

\begin{figure} 
\vskip 6.2 cm
\caption{(color-online) 
(a) Color plot of the  configuration
averaged density $\rho({\bf r})$ for a single disorder
configuration (see Fig 1(a)) from a portion of our simulation cell.
Plot (a) shows $\rho({\bf r})$ obtained from our
Monte Carlo simulations using $N = 780$ particles.
The averaging is over $1.41\times10^5$ configurations each
separated by $100$ MCS.  The average density $\rho_0
= 0.9$. Plot (b) show the same quantity obtained from 
our theory for
the corresponding disorder potential and  $\rho_0$.
} \label{f.1}
\end{figure}

A one-body random Gaussian disorder potential with
zero mean value and exponentially decaying correlations,
V$_d({\bf r})$, is constructed following a method proposed
by Chudnovsky and Dickman\cite{Chud}. The variance
($\sigma^{2}$) of the Gaussian distribution and its
spatial correlation length ($\xi$) specify the potential.
We introduce disorder with a suitably small spatial
correlation ($\xi=0.12$) in units of the interparticle
spacing.  We study the system in the fluid regime for
$\rho_{0}= 0.05,0.3$ and $0.9$, both with relatively weak
($\sigma^{2}=0.01$) and with stronger (upto $\sigma^{2}=1$)
disorder. These disorder strengths are relatively moderate
in comparison to the strength of the pair interaction,
justifying perturbation theory in the disorder potential.
We disorder average over five and ten disorder realisations
for the weak and stronger disorder cases respectively,
rejecting approximately the first $10^4 - 10^5$ Monte
Carlo Steps (MCS). We average over about $10^3 - 10^5$
configurations each separated by $10^2$ MCS, ensuring
adequate thermal averaging of the local density and
correlation functions.

For an equilibrium fluid in the absence of disorder,
all points in space are equivalent. As a consequence,
one-body distribution functions are structureless. This
property does not hold, of course, {\em within}
a particular realization of disorder. It is only
restored upon a disorder average.
If the fluid particles
do not interact, the Boltzmann relation connects the
time averaged density $\langle\rho({\bf r})\rangle$
to the local potential via $\langle\rho({\bf r})\rangle
\sim \rho_0 \exp[-\beta V_d({\bf r})]$ where $\rho_0$
is the average density of the fluid, $\beta = 1/k_B T$
with $T$ the temperature, and $V_d({\bf r})$ is the
disorder potential. This result is valid only in the
limit of vanishingly small $\rho_0$ or as $T \rightarrow
\infty$; equivalently, in the zero correlation limit.
Increasing $\rho_0$ {\em enhances} the magnitude of the
signal, making it easier to observe, but at the same time
introduces non-trivial correlations in particle positions.
Thus, in a given disorder
configuration, one-particle distributions reflect
both the inhomogeneities of the potential as well
as the consequent structuring of the local density
field as a consequence
of correlations\cite{cdgdenis}.  

Figs. 2 (a) and (b) show the time
averaged density $\rho({\bf r})$ averaged over $1.41\times10^5$
independent configurations for
$\rho_0 = .9$ and disorder strength $\sigma^2 =
0.1$. Note the strong structuring visible
in Fig 2(a). There are pronounced, correlated peaks and troughs
in the time-averaged local density, indicative both
of pinning due to local potential minima as well as
the indirect effects of inter-particle correlations.
Figs 2(b) shows the theoretically obtained plot of
$\rho({\bf r})$ using a method described below.
As can be seen, the theory captures the essential features 
of the simulation data with fair accuracy.

An inhomogeneous potential $V_{d}({\bf r})$
couples to the local number density and can
thus be absorbed into the definition of the chemical
potential $\mu$, leading to
$\mu \to \mu({\bf r}) = \mu + \delta \mu({\bf r})$
where $ \delta \mu({\bf r}) = -V_d({\bf r})$.
The Ursell function connects the local time-averaged
density $\langle\rho({\bf r})\rangle$ to the inhomogeneity
in the chemical potential via
\begin{equation}
\langle\rho({\bf r})\rangle = \langle\rho_0({\bf r})\rangle
+ \beta\int d{\bf r}^\prime S({\bf r},{\bf r}^\prime)
\delta\mu({\bf r}^\prime) + ... ,
\end{equation}
where $S({\bf r},{\bf r}^\prime)$ is defined as
$S({\bf r},{\bf r^{\prime}}) = 
\langle \rho({\bf r}) \rho({\bf r^{\prime}})\rangle - 
\langle\rho({\bf r})\rangle\langle\rho({\bf r^{\prime}})\rangle$\cite{HM}.
Results perturbative in weak disorder are obtained
by expanding about the pure limit, in which case
$\langle\rho_0({\bf r})\rangle = \rho_0$, the
average fluid density.  The density $\rho({\bf r})$
can be generated if the appropriate correlations
$\langle\rho({\bf r)}\rho({\bf r^\prime})\rangle$
and $\langle\rho({\bf r})\rangle\langle \rho({\bf
r^\prime})\rangle$ are available. To lowest order these
are the correlations of the pure system. The 
results can be extended to larger disorder 
(a non-trivial initial $\langle\rho_0({\bf r})\rangle$)
provided accurate values of $S({\bf r},{\bf r^{\prime}})$ 
computed {\em in the disordered background} are available.

For weak disorder, we may approximate
$\langle\rho({\bf r})\rho({\bf r^\prime})\rangle
- \langle\rho({\bf r}\rangle\langle\rho({\bf r^\prime})\rangle$
by $\rho_0^2 h(|{\bf r - r^\prime}|)$
where $h(r)$ is the pair correlation function of the pure system.
At somewhat stronger disorder, an alternative approach 
improves on this result by
using the {\em disorder renormalized} version of these
correlation functions, replacing $h(r)$ above by the function $g^{(1)}(r)
- g^{(2)}(r) + \delta({\bf r})/\rho_0$. Here $g^{(1)}(r)$, the disorder
averaged analog of the radial distribution function, is
\begin{equation}
g^{(1)}(r) = [\langle\rho(0)\rho(r)\rangle]/\rho_0^2 - \delta({\bf r})/\rho_0.
\end{equation}
The analog of an Edwards-Anderson parameter reflecting
the correlations of time-averaged density inhomogeneities
is the  ``off-diagonal'' distribution function $g^{(2)}(r)$,
defined through
\begin{equation}
g^{(2)}(r) = [\langle\rho(0)\rangle\langle\rho(r)\rangle]/\rho_0^2.
\end{equation}
Here, $\langle\dots\rangle\/$ denotes a thermal
average for the disordered system prior to the
disorder averaging, while the brackets $[\dots]\/$ denote an
average over disorder.

Thus, in the presence of correlations, 
the response of the density to a nonzero external
potential (apart from an overall normalization) is given by\cite{HM},
\begin{equation}
\langle\rho({\bf r})\rangle = 
\rho_{0} - \beta \rho_0^{2}  \int d{\bf r^{\prime}} 
\left(g^{(1)}(\mid{\bf r} - {\bf r^{\prime}}\mid ) 
- g^{(2)}(\mid{\bf r} - {\bf r^{\prime}}\mid ) \right)
V_{d}({\bf r^{\prime}}) - \beta \rho_0 V_{d}({\bf r}) + \ldots
\label{ursell}
\end{equation}
At this point apart from specifying how $g^{(1)}(r)$
and $g^{(2)}(r)$ are obtained\cite{gau},  which we do
below, our procedure for obtaining $\rho({\bf r})$
is complete. Using Eq.~\ref{ursell} we may obtain the 
density (Fig. 2 (b)) from the potential (Fig. 1(b)) or, by using 
a Fourier transform to invert Eq.~\ref{ursell}, 
obtain the disorder potential (Fig. 1(b)) from the 
``experimental'' density (Fig. 2(a)). In principle, inversion 
of the density to obtain $V_d({\bf r})$ requires the 
specification of $g^{(2)}(r)$ which itself depends on 
$V_d({\bf r})$. In practice, a simple iteration starting 
from pure system correlations converges rapidly. 

Our calculation of $g^{(1)}(r)$ and $g^{(2)}(r)$
is based on an early replica-based approach
to the calculation of correlation functions in
a disordered fluid\cite{gau}.  The replica method is
applied to the partition function of a system of
classical particles interacting via the Hamiltonian
$H = H_{kinetic} + \frac{1}{2}\sum_{i \ne j} V(\mid{\bf r}_i - {\bf r}_j\mid)
+ \sum_{i} V_d({\bf r}_i),$
with $V(r)$ is a two-body interaction
potential between the particles and $V_d({\bf r})$ 
a quenched, random, one-body potential
drawn from a Gaussian distribution of zero mean and short
ranged correlations: $[V_d({\bf r})V_d({\bf r^\prime})] =
K(\mid{\bf r}-{\bf r}^\prime\mid)$. One then notes that
the replicated partition function
resembles the partition 
function of a system of $n$
``species'' of particles, each labeled by an appropriate replica index,
interacting via a two-body interaction which depends both
on particle coordinates $({\bf r}_i,{\bf r}_j)$
and replica indices $(\alpha,\beta)$.
This system of $n$ species of particles is
considered to be a $n$-component
mixture~\cite{HM}.  Taking the $n \rightarrow 0$ limit in the
Ornzstein-Zernike equations governing the properties of the
mixture and assuming replica symmetry, yields the following
coupled set of equations, in Fourier space, coupling the pair correlation functions
$h^{(1)}$ and $h^{(2)}$ to the appropriate direct correlation functions
$C^{(1)}$ and $C^{(2)}$.
\begin{eqnarray}
h^{(1)}(k) & = &\frac{C^{(1)}(k) - 
[C^{(1)}(k)-C^{(2)}(k)]^{2}}{[1-C^{(1)}(k)+C^{(2)}(k)]^{2}}, \\  \nonumber
h^{(2)}(k) & = &
\frac{C^{(2)}(k)}{[1-C^{(1)}(k)+C^{(2)}(k)]^{2}}.
\label{OZ}
\end{eqnarray}

The Fourier transforms involved are defined
as $\phi(k) = \rho_0\,\int\,d{\bf r} \phi(r)\exp{(-i{\bf
k}\cdot{\bf r})}$. The replicated Ornstein-Zernike relations must be
supplemented with specific closures.  Previous work
applied to the problem of flux-lattice melting\cite{flm} 
in the presence of quenched point pinning  used
the simplest such closure, the Hyper-Netted Chain
(HNC) closure\cite{gau}. We have experimented with a variety
of closure schemes to test out the accuracy of this
approach to the calculation of fluid correlations. Our
best results are obtained with a closure scheme due
to Rogers and Young (RY)\cite{rog} for pure systems:
\begin{eqnarray}
C^{(1)}(r) & = &\exp{[-\beta(V^{(1)}(r)+V^{(2)}(r))]}\left[1 +
\frac{\exp{[Y^{(1)}(r)f(r)]}-1}{f(r)}\right] - 1 -Y^{(1)}(r), \\ \nonumber
C^{(2)}(r) & = &-\beta V^{(2)}(r),\label{rog}
\end{eqnarray}
where the function $f(r) = 1 - \exp(-\alpha\,r)$ 
interpolates between HNC and Percus-Yevick (PY)
values for large (small) $r$ and $\alpha$ the ``switching''
parameter is chosen so as to enforce thermodynamic
consistency.
Here $Y^{(\nu)}(r)\equiv h^{(\nu)}(r)-C^{(\nu)}(r) $,
$\nu = 1,2$. The pair interaction potential is
$V^{(1)}(r)$ and the off-diagonal potential $V^{(2)}(r)$
is the disorder averaged correlation function of the
disorder potential $V_{d}(r)$.  The integral equations
for the liquid state correlations as given above must
be solved numerically\cite{Gillan,flm,gau}. 

The correlation function $g^{(1)}(r)$ obtained from
self-consistent solutions of the integral
equations is qualitatively similar to that for the
pure case at low and intermediate levels of disorder.
At large values of disorder, a disorder-induced
suppression of structure in $g^{(1)}(r)$ is clearly
apparent. This is shown in Fig. 3(a) which exhibits
the quantity $\Delta g(r) = g^{(1)}(r) - g(r)$, the
difference between the pair distribution function
in the disordered and pure cases. A
direct comparison to the simulation data is 
shown in Fig. 3 (b). We find that while HNC and
PY respectively under and over estimate\cite{HM}
correlations, the RY closure gives very good agreement
with our simulations. (This is consistent with results
for the pure system.) The parameter $\alpha$ in the
latter is taken to be its pure system value
of $0.3$, a value which agrees best with 
simulation data over the full range of densities and
disorder strengths.

\begin{figure}
\includegraphics[height=6cm]{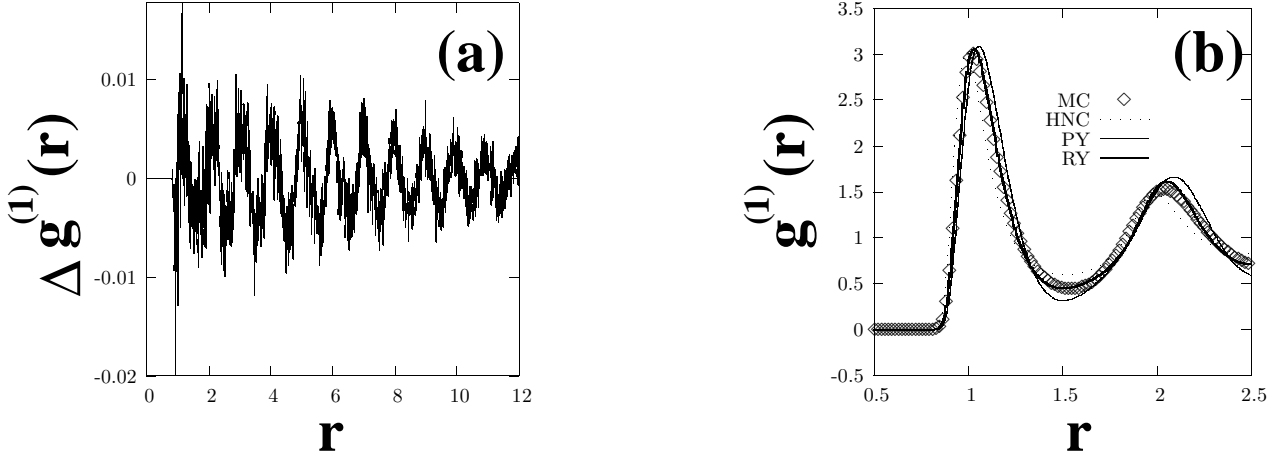}
\caption{(a) Plot of $\Delta g^{(1)}(r)$ for a single density $\rho_0 =
0.9$ and $\sigma^2 = 1.0$. (b) Comparison of $g^{(1)}(r)$ for density $\rho_0 =
0.9$ and $\sigma^2 = 1.0$ of simulation data (MC) with the
results of HNC, PY and RY closures. Ten disorder configurations were 
used in the averaging.}
\label{f.2}
\end{figure}

\begin{figure}
\includegraphics[height=6cm]{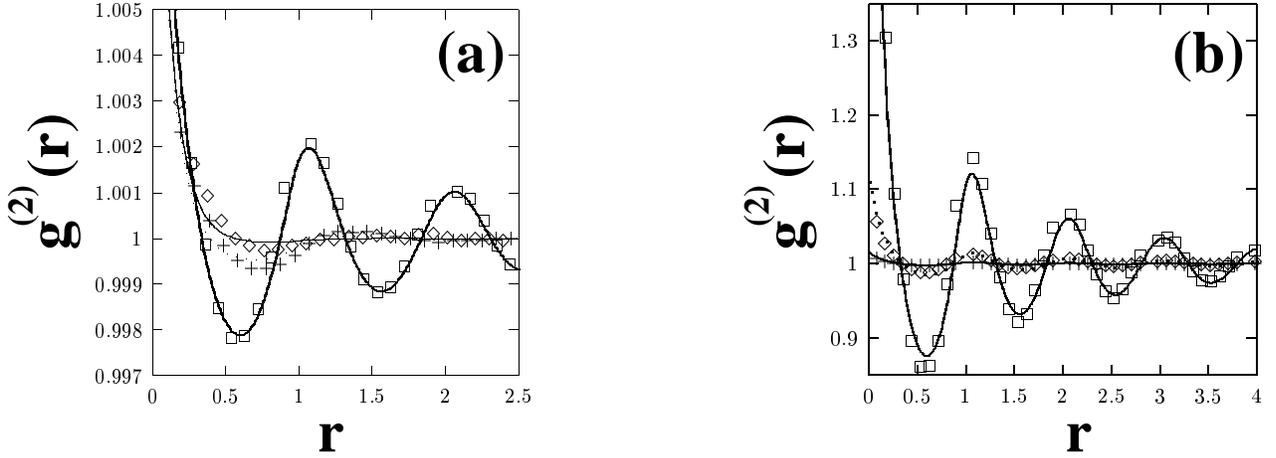}
\caption{ (a) Plot of $g^{(2)}(r)$ comparing simulations (points)
and result from the RY closure (lines) for densities $\rho_0 =
.05\,\,(+), .3$($\diamond$) and $.9$(box) for $\sigma^2 =
.01$. Ten disorder configurations were used in the
averaging. (b) Plot of $g^{(2)}(r)$ -- simulations and
theory (RY) for $\sigma^2 = .01\,\, (+), .1$\,\,($\diamond$)
and $1$ (box) for $\rho_0 = 0.9$ }
\label{f.3}
\end{figure}

We show the off-diagonal correlation function $g^{(2)}(r)$ at
$\sigma^2 = 0.01$ for $\rho_0 = .05, .3$ and $.9$ in
Fig. 4(a). In the simulations, this correlation function
is computed by extracting the correlations of the local
time-averaged density in a particular configuration
of disorder and then averaging over several disorder
realizations. Correlations increase with increasing
density as expected, as manifest in the oscillations
of $g^{(2)}(r)$. For the same reason, the off-diagonal
correlations for any density also increase with the
strength of the disorder correlator (Fig. 4(b)).  In both
cases, our theoretical estimates based on the RY closure
agree to good accuracy with correlation functions obtained
from the simulations.

Liquid state theory based approaches are easily adapted
to a variety of different problems {\it vis a vis.}
direct simulations. Thus, the approach presented here
is potentially useful to a variety of theoretical models
for experimental data.  The following limitations must,
however, be kept in mind: First, $\rho({\bf r})$ must
be accurately determined from experimental pictures
for correlations to be obtained correctly. Since
the data are often thresholded, reflecting some level
of coarse graining of the amplitude of the signal, a
direct comparison with theory may not always be 
straighforward. (Interestingly, the size of the
probe particles {\em do not} impose any theoretical
limit\cite{note} on the spatial resolution of $V_d({\bf
r})$ -- this is limited only by the accuracy with which
$\rho({\bf r})$ can be imaged.) Second, our analysis
neglects higher order density-density correlations.
Finally, at high disorder strengths and strong
correlations, the possibility exists of a transition into
a state in which replica symmetry is broken, a transition
which would not be captured in this calculation\cite{thal}.

What types of experiments might access the unusual
``off-diagonal'' correlation function $g^{(2)}$ ?
Any experiment in which repeated `snapshots'' of
the system are taken, with sufficient statistics,
should yield results which can be addressed by
these methods.  For vortices in superconductors,
scanning tunneling spectroscopy\cite{hess,eskildsen}
magneto-optic imaging\cite{pal} and Lorentz and
magnetic force microscopy\cite{grier,volodin} can all
be used as structural probes of vortex structure in
thin superconducting films.  Magnetic bubble domain
arrays are another experimental system in which many
of the features described here should be accessible
\cite{seshadri1}. We also note that several recent
experiments directly access the configurations of a large
number of colloidal particles confined to two dimensions
\cite{cherry,larsen,zahn,korda}.
In these experiments, extensive statistics for particle
positions can be generated.  The colloid literature
has hitherto concentrated on making substrates as
homogeneous and disorder-free as possible. However,
as we argue here, interacting colloidal particles moving on
disordered substrates can exhibit interesting and
non-trivial correlations which have not hitherto been
characterized. Experimental work in this direction would
be welcome.

{\bf Acknowledgements:} Discussions with C. Dasgupta, S. Sastry
and M. Rao and financial support from DST grant SP/S2/M-20/2001 
are gratefully acknowledged.

\end{document}